\newif\ifanonymous
\newif\ifdraft

\anonymousfalse
\draftfalse
\InputIfFileExists{localflags}{}

\documentclass[sigconf]{acmart}

\usepackage{versions}

\excludeversion{conf}
\includeversion{full}

\usepackage{wasysym}
\usepackage{stmaryrd}
\usepackage{multirow}
\usepackage{float}
\usepackage{graphicx}
\usepackage{times}
\usepackage{helvet}
\usepackage{courier}
\usepackage{enumerate}
\usepackage{paralist}
\usepackage{url,etoolbox}
\usepackage[normalem]{ulem}
\usepackage{pgf}
\usepackage{tikz}
\usepackage[utf8]{inputenc}
\usetikzlibrary{arrows,shapes,patterns,positioning,fit}
\usepackage{thmtools,thm-restate}

\tikzset{
    every node/.style = {
    align=center
    },
    icon/.style= {
        },
    tool/.style= {
         shape=rectangle, draw, minimum height=1.6em,
    rounded corners,
        },
    file/.style= {
         shape=rectangle, draw, minimum height=1.3em,
         rounded corners,
         font=\ttfamily,
         fill=white,
        },
    user/.style= {
         shape=rectangle, draw, minimum height=1.3em,
    rounded corners,
        },
    dist/.style= {
         shape=circle, draw, minimum height=1.6em,
    rounded corners,
        },
    var/.style= {
        },
}

\usepackage[vlined]{algorithm2e} 
\ifx \DontPrintSemicolon \undefined
 \def \DontPrintSemicolon {\dontprintsemicolon}
\fi

\makeatletter
\DeclareRobustCommand{\defeq}{\mathrel{\rlap{%
  \raisebox{0.3ex}{$\m@th\cdot$}}%
  \raisebox{-0.3ex}{$\m@th\cdot$}}%
  =}
\DeclareRobustCommand{\eqdef}{=\mathrel{\rlap{%
  \raisebox{0.3ex}{$\m@th\cdot$}}%
  \raisebox{-0.3ex}{$\m@th\cdot$}}%
  }
\makeatother
\definecolor{ToDoColor}{rgb}{0,0.16,0.90} 
\definecolor{OutlineColor}{rgb}{0.2,0.8,0.2} 
\definecolor{CommentColor}{rgb}{0.90,0.16,0} 

\def \R {\mathbb R}

\newcommand{\commentout}[1]{}

\newcommand{\ie}{i.\,e.}
\newcommand{\eg}{e.\,g.}
\newcommand{\cf}{cf.}

\newcommand{\net}{\ensuremath{\mathsf{N}}}
\newcommand{\ptask}{\ensuremath{\Pi}}
\newcommand{\mtask}{\ensuremath{\mathsf{M}}}
\newcommand{\fixes}{{\ensuremath{\mathsf{F}}}}
\newcommand{\attacks}{{\ensuremath{\mathsf{A}}}}

\newcommand{\props}{\ensuremath{\mathsf{P}}}

\newcommand{\init}{\ensuremath{\mathsf{I}}}
\newcommand{\goal}{\ensuremath{\mathsf{G}}}

\newcommand{\pre}{\ensuremath{\mathit{pre}}}
\newcommand{\post}{\ensuremath{\mathit{post}}}

\newcommand{\nprops}{\ensuremath{\props^\net}}
\newcommand{\ninit}{\ensuremath{\init^\net}}
\newcommand{\mbudget}{\ensuremath{b^\mtask}}

\newcommand{\aprops}{\ensuremath{\props^\attacks}}
\newcommand{\ainit}{\ensuremath{\init^\attacks}}

\newcommand{\abudget}{\ensuremath{b^\attacks}}

\newcommand{\minmbudget}{\ensuremath{\mbudget_\mathit{min}}}
\newcommand{\minabudget}{\ensuremath{\abudget_\mathit{min}}}
\newcommand{\mbudgetfactor}{\ensuremath{\gamma_\mtask}}
\newcommand{\abudgetfactor}{\ensuremath{\gamma_\attacks}}

\newcommand{\mitigation}{\ensuremath{\sigma}}
\newcommand{\pareto}{{\ensuremath{\mathcal{F}}}}

\newcommand{\outs}{\ensuremath{O}}
\newcommand{\apply}[1]{\ensuremath{\llbracket #1 \rrbracket}}

\newcommand{\app}{\mathit{app}}

\newcommand{\subnet}{\mathsf{subnet}}
\newcommand{\haclz}{\mathsf{haclz}}
\newcommand{\vulexists}{\mathsf{vul\_exists}}

\newcommand{\controls}{\mathsf{controls}}
\newcommand{\available}{\mathsf{available}}
\newcommand{\compromised}{\mathsf{compromised}}
\newcommand{\targetundercontrol}{\mathsf{zcompromised}}
\newcommand*\integrity{%
  \begingroup\mathsf{\xdef\tmp{\fam\the\fam\relax}}\tmp
  in\-te\-gri\-ty\endgroup}
\newcommand*\confidentiality{%
  \begingroup\mathsf{\xdef\tmp{\fam\the\fam\relax}}\tmp
  con\-fi\-dent\-ia\-lity\endgroup}
\newcommand*\availability{%
  \begingroup\mathsf{\xdef\tmp{\fam\the\fam\relax}}\tmp
  a\-vail\-abi\-lity\endgroup}
\newcommand{\fwapplied}{\mathsf{fwapplied}}

\newcommand{\randomvariable}[1]{\mathit{#1}}

\newcommand{\rHosts}{\randomvariable{H}}
\newcommand{\rConfs}{\randomvariable{C}}
\newcommand{\rUnif}{\randomvariable{U}}
\newcommand{\rLength}{\randomvariable{N}}
\newcommand{\rVuln}{\randomvariable{V}}

\newcommand{\NVD}{\mathit{NVD}}

\newcommand{\port}{\mathit{port}}
\newcommand{\proto}{\mathit{proto}}
\newcommand{\cve}{\mathit{cve}}

\copyrightyear{2019}
\acmYear{2019}
\setcopyright{acmlicensed}
\acmConference[SAC'19]{ACM SAC Conference}{April 8-12, 2019}{Limassol, Cyprus}
\acmBooktitle{The 34th ACM/SIGAPP Symposium on Applied Computing (SAC '19), April 8--12, 2019, Limassol, Cyprus}
\acmArticle{4}
\acmPrice{15.00}
\acmDOI{10.1145/3297280.3297473}
\acmISBN{978-1-4503-5933-7/19/04}

\begin{document}

\begin{conf}
\title{Towards Automated Network Mitigation Analysis}
\end{conf}

\begin{full}
\title{Towards Automated Network Mitigation Analysis (extended)}
\end{full}

\ifanonymous\else






\author{
Patrick Speicher\textsuperscript{*}, Marcel Steinmetz\textsuperscript{*}, J\"org Hoffmann\textsuperscript{$\dagger$}, Michael Backes\textsuperscript{*}, and Robert K\"unnemann\textsuperscript{*}
}
\affiliation{
\vspace{-0.4cm}
   \textsuperscript{*}CISPA Helmholtz Center\\
   \textsuperscript{$\dagger$}CISPA Helmholtz Center, Saarland University\\
   Email: \{first name\}.\{last name\}@cispa.saarland,  backes@cispa.saarland, hoffmann@cs.uni-saarland.de\\
}

\renewcommand{\shortauthors}{Speicher et al.}

\fi

\renewcommand{\textrightarrow}{$\rightarrow$}

\begin{abstract}
    Penetration testing is a well-established practical concept for
    the identification of potentially exploitable security weaknesses
    and an important component of a security audit. 
    Providing a holistic security assessment for networks
    consisting of several hundreds hosts is hardly feasible though without
    some sort of mechanization. 
    Mitigation, prioritizing counter-measures subject to a given budget,
    currently lacks a solid theoretical understanding and is hence more
    art than science.
    In this work, we propose the first approach for
    conducting comprehensive what-if analyses in order to reason about
    mitigation in a conceptually well-founded manner.
    To evaluate and compare mitigation strategies,
    we use \emph{simulated penetration testing},
    i.e., automated attack-finding, based on a network model to which
    a subset of a
    given set of mitigation actions, e.g., changes to the network
    topology, system updates, configuration changes etc.\ is applied.
    Using \emph{Stackelberg planning}, we determine optimal
    combinations that minimize the maximal attacker success (similar
    to a Stackelberg game), and thus provide a well-founded basis for
    a holistic mitigation strategy.
    We show that these Stackelberg planning models can largely be
    derived from network scan, public vulnerability databases and
    manual inspection with various degrees of automation and detail,
    and we simulate mitigation analysis on networks of different size
    and vulnerability.
    \end{abstract}



%

\maketitle

\section*{Notes about this version}

The first version of this article was published on arXiv
under the title
`Simulated Penetration Testing and Mitigation Analysis'~\cite{DBLP:journals/corr/00010KSS17v1}.
The mitigation analysis formalism was later dubbed `Stackelberg
planning' and discussed in a more general scope in a separate
publication~\cite{aaai-stackelberg}.
The present version thus concentrates on the application to simulated
pentesting. In comparison to the
previous version, the algorithmic implementation was removed (it can
be found~\cite{aaai-stackelberg}), the presentation was streamlined,
typos were fixed and the title changed to reflect the new focus.

\section{Introduction}

    Penetration testing (pentesting) 
    evaluates the security of an IT infrastructure by trying to identify and 
    exploit vulnerabilities. 
    It constitutes
    a central, often mandatory component of a security audit,  e.g.,
    the Payment Card Industry Data Security Standard prescribes
    `network vulnerability scans at least quarterly and
    after any significant change in the network'~\cite{pcidss}.
    Network pentests are frequently conducted on networks with
    hundreds of machines. 
    Here, the vulnerability of the network is
    a combination of
    host-specific weaknesses that compose to an attack.
    Consequently, an exhausting search is out of question,
    as
    the search space for these combinations grows exponentially 
    with the number of hosts.
    Choosing the right attack vector 
    requires a vast amount of experience, arguably making network pentesting more art than science.
    
    While it is conceivable that an experienced analyst
    comes up with several of the most severe attack vectors,
    this is not sufficient to provide for a sound 
    mitigation strategy,
    as the evaluation of a mitigation strategy 
    requires a holistic security assessment.
    So far, there is no rigorous foundation for what is 
    arguably the most important step, the step \emph{after} the
    pentest: how to
    mitigate these vulnerabilities.

    In practice, the severity of weaknesses is assessed more or less
    in isolation, proposed counter-measures all too often focus on single
    vulnerabilities, and the mitigation path is left to the customer.
    There are exceptions, but they require considerable manual effort.

    \emph{Simulated pentesting} was proposed to
    automate large-scale network testing by simulating the attack finding process based
    on a logical model of the network. 
    The model may be generated from network scans, public
    vulnerability databases and manual inspection with various degrees
    of automation and detail.
    To this end, AI planning methods have been proposed
    \cite{boddy:etal:icaps-05,lucangeli:etal:secart-10} and in fact
    used commercially, at a company called Core Security,
    since at least
    2010~\cite{core-impact}.
    These approaches, 
    which derive from earlier approaches based on attack graphs \cite{philipps:swiler:nsp-98,schneier:dobbs-99,sheyner:etal:ssp-02},
    assume complete knowledge over the network
    configuration, which is often unavailable to the modeller, as well
    as the attacker. 
    We follow a more recent
    approach favouring 
    Markov decisions processes (MDP) as the underlying state model to
    obtain a good middle ground between accuracy and practicality~\cite{durkota:lisy:stairs-14,hoffmann:icaps-15} (we discuss this in detail as part of our related work discussion, Section~\ref{sec:related-work}).

    Simulated pentesting has been used to great success,
    but an important feature was overseen so far. If a model of the
    network is given, one can reason about possible mitigations
    without implementing them -- namely, by simulating the attacker on a modified
    model.
    This allows for analysing and comparing different mitigation
    strategies in terms of the (hypothetical) network resulting from
    their application.
    This problem was recently introduced as \emph{Stackelberg planning} in the
    AI community~\cite{aaai-stackelberg}.
    Algorithmically, the attacker-planning problem now becomes part of a larger
    what-if planning problem, in which the best mitigation plans are constructed.
    \begin{full}
    This min-max notion is similar to a Stackelberg game,
    which are frequently used in security games~\cite{Korzhyk:2011:SVN:2051237.2051246}.
    The foundational assumption is that the 
    defender acts first, while the adversary can choose her best
    response after observing this choice, similar to a market leader
    and her followers.
    The algorithm thus provides a well-founded basis for a holistic
    mitigation strategy.
\end{full}

    Mitigation actions can represent, but are not limited to,
    changes to the network topology, e.g., adding a packet filter,
    system updates that remove vulnerabilities,
    and configuration changes or application-level firewalls which
    work around issues.
    \begin{full}
    While, e.g., an application-level firewall might be an
    efficient temporary workaround for a vulnerability that affects a single host,
    contracting a software vendor to provide a patch might be more
    cost-efficient in case the vulnerability appears throughout the network.
    To reflect cases like this,
    mitigation actions are assigned a cost for their first application (set-up cost),
    and another potentially different cost for all subsequent applications (application cost).
    \end{full}
    The algorithm computes optimal combinations w.r.t.\  minimizing
    the maximal attacker success for a given budget, and proposes
    dominant mitigation strategies with respect to cost and attacker
    success probability.

    After discussing related work in Section~\ref{sec:related-work}
    and giving a running example in Section~\ref{sec:running-example},
    we present the mitigation analysis model 
    in
    Section~\ref{sec:formalism}, framed in
    a formalism suited for a large range of mitigation/attack planning
    problems.
    In Section~\ref{sec:model-acquisition}, we show how to derive
    these models by scanning a given
    network using the Nessus network-vulnerability scanner.
    The attacker action model is then derived using
    a vulnerability database and data associated using the Common Vulnerability
    Scoring System (CVSS).
    This methodology provides a largely automated method of deriving
    a model (only the network topology needs to be given by hand),
    which can then be used as it is, or further refined.
    In Section~\ref{sec:experiments}, we evaluate our algorithms w.r.t.\ 
    problems from this class, derived from a vulnerability database
    and a simple scalable network topology.

\section{Related Work}\label{sec:related-work}

Our work is rooted in a long line of research on network security
modeling and analysis, starting with the consideration of \emph{attack
  graphs}. The simulated pentesting branch of this research
essentially formulates attack graphs in terms of standard sequential
decision making models --- \emph{attack planning} --- from AI\@. We
give a brief background on the latter first, before considering the
history of attack graph models.

Automated Planning is one of the oldest sub-areas of AI
(see~\cite{ghallab:etal:04} for a comprehensive introduction). The
area is concerned with general-purpose planning mechanisms that
automatically find a \emph{plan}, when given as input a high-level
description of the relevant world properties (the \emph{state
  variables}), the \emph{initial state}, a \emph{goal} condition, and
a set of \emph{actions}, where each action is described in terms of a
\emph{precondition} and a \emph{postcondition} over state variable
values. In \emph{classical planning}, the initial state is completely
known and the actions are deterministic, so the underlying state model
is a directed graph (the \emph{state space}) and the plan is a path
from the initial state to a goal state in that graph. In
\emph{probabilistic planning}, the initial state is completely known
but the action outcomes are probabilistic, so the underlying state
model is a Markov decision process (MDP) and the plan is an action
\emph{policy} mapping states to actions.\begin{full}%
In \emph{partially observable
  probabilistic planning}, we are in addition given a probability
distribution over the possible initial states, so the underlying state
model is a partially observable MDP (POMDP). 
\end{full}

The founding motivation for Automated Planning mechanisms is flexible
decision taking in autonomous systems, yet the generality of the
models considered lends itself to applications as diverse as the control
of modular printers~\cite{ruml:etal:jair-11}, natural language
sentence generation~\cite{koller:hoffmann:icaps-10,koller:petrick:ci-11}, %
\begin{full}%
greenhouse
logistics~\cite{helmert:lasinger:icaps-10}, 
\end{full}%
 and, in particular,
network security penetration testing~\cite{boddy:etal:icaps-05,lucangeli:etal:secart-10,sarraute:etal:aaai-12,durkota:lisy:stairs-14,hoffmann:icaps-15}. 
\begin{full}
    This
latter branch of research --- network attack planning as a tool for
automated security testing --- has been coined simulated pentesting,
and is what we continue here.
\end{full}

Simulated pentesting is rooted in the consideration of attack graphs,
first introduced by Philipps and Swiler~\cite{philipps:swiler:nsp-98}. An attack graph breaks down the space
of possible attacks into atomic components, often referred to as
attack actions, where each action is described by a conjunctive
precondition and postcondition over relevant properties of the system
under attack. This is closely related to the syntax of classical
planning formalisms. Furthermore, the attack graph is intended as an
analysis of threats that arise through the possible
\emph{combinations} of these actions. This is, again, much as in
classical planning. That said, attack graphs come in many different
variants, and the term ``attack graph'' is rather overloaded. From our
point of view here, relevant lines of distinction are the following.

In several early works
(\eg\ \cite{schneier:dobbs-99,templeton:levitt:nspw-00}), the attack
graph is the attack-action model itself, presented to the human as an
abstracted overview of (atomic) threats. It was then proposed to
instead reason about combinations of atomic threats, where the attack
graph (also: ``full'' attack graph) is the state space arising from
all possible sequencings of attack actions
(\eg\ \cite{ritchey:amman:ssp-00,sheyner:etal:ssp-02}). Later,
positive formulations --- positive preconditions and postconditions
only --- where suggested as a relevant special case, where attackers
keep gaining new assets, but never lose any assets during the course
of the attack~\cite{templeton:levitt:nspw-00,ammann:etal:ccs-02,jajodia:etal:ctiac-05,ou:etal:ccs-06,noel:etal:catch-09,ghosh:ghosh:secart-09}. This
restriction drastically simplifies the computational problem of
non-probabilistic attack graph analysis, yet it also limits expressive
power, especially in probabilistic models where a stochastic effect of
an attack action (\eg, crashing a machine) may be detrimental to the
    attacker's objectives.\begin{full}\footnote{The restriction to positive
  preconditions and postconditions is actually known in Automated
  Planning not as a planning problem of interest in its own right, but
  as a problem \emph{relaxation}, serving for the estimation of goal
  distance to guide search on the actual problem~\cite{bonet:geffner:ai-01,hoffmann:nebel:jair-01}.}\end{full}

\begin{full}
A close relative of attack graphs are \emph{attack trees}
(\eg\ \cite{schneier:dobbs-99,mauw:oostdijk:icisc-05}). These arose
from early attack graph variants, and developed into `Graphical
Security Models'~\cite{kordy:etal:qest-13}: Directed acyclic AND/OR
graphs organizing known possible attacks into a top-down refinement
hierarchy. The human user writes that hierarchy, and the computer
analyzes how attack costs and probabilities propagate through the
hierarchy. In comparison to attack graphs and planning formulations,
this has computational advantages, but cannot find unexpected attacks,
arising from unforeseen combinations of atomic actions.
\end{full}

Probabilistic models of attack graphs/trees have been considered
widely
(\eg\ \cite{buldas:etal:critis-06,niitsoo:iwsec-10,sarraute:etal:aisec-11,singhal:ou:nist-11,buldas:stepanenko:gamesec-12,lisy:pibil:paisi-13,homer:etal:jcs-13}),
\begin{full}
yet they weren't, at first, given a formal semantics in terms of
standard sequential decision making formalisms. The latter was done
later on by the AI community in the simulated pentesting branch of
research. After initial works linking non-probabilistic attack graphs
to classical planning~\cite{boddy:etal:icaps-05,lucangeli:etal:secart-10}, Sarraute et
al.~\cite{sarraute:etal:aaai-12} devised a comprehensive model based
on POMDPs, designed to capture penetration testing as precisely as
possible, explicitly modeling the incomplete knowledge on the
attacker's side, as well as the development of that knowledge during
the attack. As POMDPs do not scale --- neither in terms of modeling nor
in terms of computation --- it was thereafter proposed to use MDPs as a
more scalable intermediate model~\cite{durkota:lisy:stairs-14,hoffmann:icaps-15}. 
\end{full}
\begin{conf}
    and were later linked to  to classical planning~\cite{boddy:etal:icaps-05,lucangeli:etal:secart-10}.
More precise formulations in terms of 
    \emph{partially observable MDPs}
were proposed for their ability to model incomplete knowledge on the
    attacker's side~\cite{sarraute:etal:aaai-12}.
As POMDPs do not scale --- neither in terms of modeling nor
in terms of computation --- it was thereafter proposed to use MDPs as a
more scalable intermediate model~\cite{durkota:lisy:stairs-14,hoffmann:icaps-15}.
\end{conf}
Here we build upon this latter model.

Stackelberg planning~\cite{aaai-stackelberg}
models not only the attacker, but also the
defender, and in that sense relates to more general game-theoretic
security models. 
The most prominent application of such models thus far
concerns physical infrastructures and defenses (\eg\ \cite{tambe:11}),
quite different from the network security setting.
A line of research
considers attack-defense trees
(\eg~\cite{kordy:etal:fast-10,kordy:etal:qest-13}), not based on
standard sequential decision making formalisms. Some research
considers pentesting but from an abstract theoretical perspective~\cite{boehme:felegyhazi:gamesec-10}. A basic difference to most
game-theoretic models is that our mitigation analysis does not
consider arbitrarily long exchanges of action and counter-action, but
only a single such exchange: defender applies network fixes, attacker
attacks the fixed network. 
\begin{full}
The latter relates to Stackelberg
competitions, yet with interacting state-space search models
underlying each side of the game.
\end{full}

\section{Running Example}\label{sec:running-example}

\begin{figure}
\centering
\includegraphics[width=\linewidth]{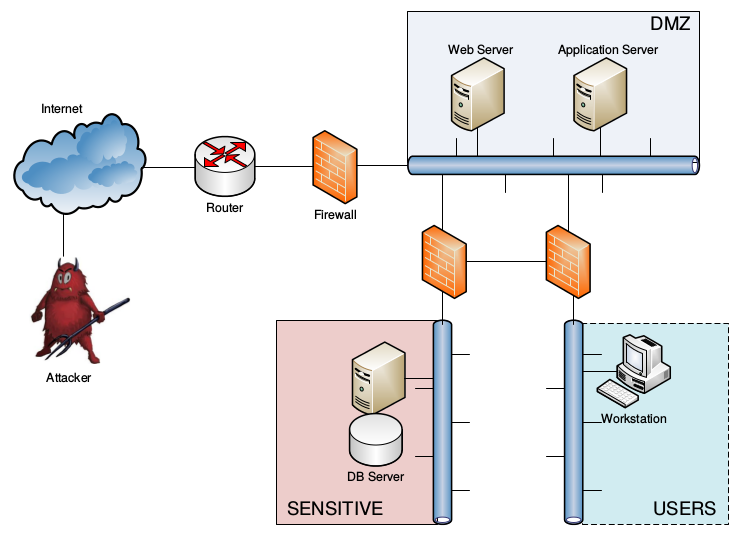}
\vspace{-0.8cm}
\caption{Network structure in our running example. (Figure adapted
  from Sarraute et al.\ \cite{sarraute:etal:aaai-12}.)}\label{fig:running-example-}
  \vspace{-0.8cm}
\end{figure} 

We will use the following running example
for easier introduction of our formalism 
and 
to foreshadow the modelling of networks
which we will use in Section~\ref{sec:model-acquisition}.
Let us
consider a network of five hosts, i.e., computers that are assigned an
address at the network layer.
It consists of 
a webserver $W$,
an application server $A$,
a database server $D$,
and a workstation $S$.
We partition the network into three zones called as follows:
\begin{inparaenum}[1)]
    \item the sensitive zone, which contains important assets, i.e.,
        the database server $D$ 
    \item the DMZ, which contains the services that need to be
        available from the outside, i.e., $A$ and $W$,
    \item the user zone, in which $S$ is placed and
    \item the internet, which is assumed under adversarial control by
        default and contains at least a host $I$.
\end{inparaenum}

These zones are later (cf. Section~\ref{sec:experiments}) used
to define the adversarial goals and may consist of several subnets.
For now, 
each zone except the internet consists of exactly one subnet.
These subnets are interconnected, with the exception of the
internet, which is only connected to the DMZ\@. 
Firewalls filter some packets transmitted between the zones. We will
assume that the webserver can be accessed via HTTPS (port 443) from the internet.

\section{Mitigation analysis as Stackelberg planning}\label{sec:formalism}

It was recently proposed to model penetration testing and mitigation
tasks as Stackelberg planning task~\cite{aaai-stackelberg}. We review
this formalism and show how vulnerability analysis can be mapped onto
it.

Intuitively, the attacks we consider might make a service unavailable,
but not physically remove a host from the network or
add a physical connection between two hosts.
We thus 
distinguish between network propositions and attacker propositions,
where the former describes the network infrastructure and 
persistent configuration,
while the latter describes the attacker's advance through the network.
By means of this distinction, we may
assume the state of the network to be fixed,
while everything else can be manipulated by the attacker.
The network state will, however,
be altered during mitigation
analysis, which we will discuss in more detail afterwards.

Networks are logically described through a finite set of \emph{network
propositions} $\nprops$. A concrete \emph{network state} is a subset of
network propositions $s^\net \subseteq \nprops$ that are true in this state.
All propositions $p \not \in s^\net$ are considered to be false.

\begin{example}\label{ex:network-topology}
    In the running example, 
    the network topology is described in terms of
    network propositions 
    $\subnet(s,h) \in \nprops$ assigning a host $h$ to a subnet $s$,
    e.g., $\subnet(\mathsf{sensitive},D)\in\nprops$.
    Connectivity is defined between subnets, e.g.,
    $\haclz(\mathsf{internet},\mathsf{dmz},443,\mathsf{tcp})\in\nprops$
    indicates that
    TCP packets 
    with destination port $443$ (HTTPS) can pass from
    the internet into the DMZ.
    We assume that the webserver $W$, the workstation $S$ and the
    database server $D$ are vulnerable, e.g.,
    $\vulexists(\cve_W,W,443,\mathsf{tcp},\integrity)\in\nprops$
    for a vulnerability with CVE identifier $\cve_W$ affecting
    $W$ on TCP port 443, that compromises integrity.
\end{example}

We formalize network penetration tests in terms of a probabilistic planning problem:

\begin{definition}[penetration testing task~\cite{aaai-stackelberg}]\label{def:task}
A \emph{penetration testing task} is a tuple $\Pi = (\aprops, \attacks,
\ainit, \goal, \abudget_0)$ consisting of:
    \begin{itemize}
        \item a finite set of \emph{attacker propositions} $\aprops$,
        \item a finite set of (probabilistic) \emph{attacker actions} $\attacks$ (cf.
            Definition~\ref{def:attacker-acts}),
        \item the attacker's \emph{initial state} $\ainit\subseteq \aprops$,
        \item  a conjunction $\goal$ over attacker proposition
literals, called the \emph{attacker goal}, and
        \item  a non-negative attacker \emph{budget} $\abudget \in \R^+ \cup
            \{\infty\}$,
            including the special case of an unlimited budget $\abudget=\infty$.
    \end{itemize}
\end{definition}

The objective in solving such a task --- the attacker's objective ---
will be to maximize attack probability, \ie, to find action strategies
maximizing the likelihood of reaching the goal, which we will specify
in more detail.
The attacker proposition are used to describe the state of the attack, \eg,
dynamic aspects of the network and which
hosts the attacker has gained access to.
\begin{example}
    Consider an attacker that initially controls the internet, i.e.,
    $\controls(I)\in\ainit$
    and has not yet caused $W$ to crash,
    $\available(W)\in\ainit$.
    The attacker's aim might be to inflict a privacy-loss on $D$,
    i.e.,
    $\compromised(D,\mathsf{privacy})$,
    with
    a budget $\abudget$ of $3$ units, which relate to the attacker actions below.
\end{example}

The attacks themselves are described in terms of actions which can
depend on 
both network and attacker propositions, but only influence the attacker state.

\begin{definition}[attacker actions~\cite{aaai-stackelberg}]\label{def:attacker-acts}
An \emph{attacker action} $a \in \attacks$ is a tuple 
    $(\pre^\net(a), \pre^\attacks(a), c(a), \outs(a))$
where
\begin{itemize}
   \item $\pre^\net(a)$ 
       is a conjunction over network proposition literals
        called the \emph{network-state precondition},
   \item $\pre^\attacks(a)$
       is a conjunction over attacker proposition literals
       called  the \emph{attacker-state precondition},
   \item $c(a) \in \R^+$ is the \emph{action cost}, and
   \item $\outs(a)$ is a finite set of \emph{outcomes}, 
       each $o\in\outs(a)$
       consisting of an \emph{outcome probability} 
        $p(o) \in (0, 1]$ 
        and a \emph{postcondition}
        $\post(o)$
        over attacker proposition literals.
        We assume that $\sum_{o \in \outs(a)} p(o) = 1$.
\end{itemize}
\end{definition}

The stochastic effect $\post(o) \in \outs(a)$ can be used to model attacks that are probabilistic by nature, as well as to model incomplete knowledge (on the attacker's side) about
the actual network configuration. Because $\post(o)$ is limited to attacker propositions, we implicitly assume
that the attacker cannot have a direct influence on the network itself. Although
this is restrictive, it is a common assumption in the penetration testing
literature (\eg~\cite{jajodia:etal:ctiac-05,ou:etal:ccs-06,noel:etal:catch-09,ghosh:ghosh:secart-09}).
The attacker action cost can be used to represent the effort the attacker has to
put into executing what is being abstracted by the action. This can, e.g.,
be the estimated amount of time an action requires to be carried out, or the
actual cost in terms of monetary expenses.

\begin{example}\label{ex:acts}
    If an attacker controls a host which can access a second host
    that runs a vulnerable service, it
    can compromise the second host w.r.t.\ privacy,
    integrity or availability, depending on the vulnerability.
    This is reflected, e.g., by an attacker action $a\in\attacks$ 
    which requires
    access to a vulnerable $W$ within the DMZ,
    via the internet, s.t. $
        \pre^\net(a) =   \subnet(\mathsf{dmz},W) \land   \subnet(\mathsf{internet},I)
         \land \haclz(\mathsf{internet},\mathsf{dmz},443,\mathsf{tcp}) 
         \land \vulexists(\cve_W,W,443,\mathsf{tcp},\integrity) $.
    In addition,  $I$ needs to be under adversarial
    control (which is the case initially), and $W$ be available:
    $\pre^\attacks(a)=\controls(I)\land \available(W)$.

    The cost of this known vulnerability may be set to $c(a)=1$,
    in which case
    the
    adversarial budget above relates to the number of such
    vulnerabilities used. More elaborate models are
    possible to distinguish known vulnerabilities from
    zero-day exploits which may exists, but only be
    bought or developed at high cost, or threats arising from social
    engineering.
    
    We define three different outcomes
    $O(a)=\{o_\mathit{success},\allowbreak o_\mathit{fail},\allowbreak
    o_\mathit{crash}\}$ with probabilities
    \begin{itemize}
        \item 
    $\post(o_\mathit{success}) =\allowbreak \compromised(W, \integrity) \allowbreak \land
    \controls(W)$ in case the exploit succeeds,
\item
    $\post(o_\mathit{fail})=\top$ in case the exploit has no effect
            and
\item     and $\post(o_\mathit{crash})=\neg \available(W)$ if it crashes
    $W$. 
    \end{itemize}
For example, we may have $p(o_\mathit{success}) = 0.5$,
    $p(o_\mathit{fail}) = 0.49$, and $p(o_\mathit{crash}) = 0.01$
    because the exploit is of stochastic nature, with a small 
    probability to crash the machine.

    Regarding the first action outcome, $o_\mathit{success}$, note
    that we step here from a vulnerability that affects integrity, to
    the adversary gaining control over $W$. This is, of course, not a
    requirement of our formalism; it is a practical design decision
    that we make in our current model acquisition setup (and that was
    made by previous works on attack graphs with similar model
    acquisition machinery
    \eg\ \cite{ou:etal:ccs-06,singhal:ou:nist-11}), because the
    vulnerability databases available do not distinguish between a
    privilege escalation and other forms of integrity violation. We
    get back to this in Section~\ref{sec:model-acquisition}.
    Regarding the third action outcome, $o_\mathit{crash}$, note that
    negation is used to denote removal of literals, \ie, the following
    attacker state will not contain $\available(W)$ anymore, so that
    all vulnerabilities on $W$ cease to be useful to the attacker.
\end{example}

The syntax and state transition semantics just specified is standard
probabilistic planning. Thus, the state space of a penetration testing task 
can be viewed as a Markov decision process (MDP). A solution for an MDP
is called policy and there are various objectives for these policies, \ie, notions of
optimality, in the literature. For attack planning, arguably the most
natural objective is \emph{success probability}: the likelihood that
the attack policy will reach a goal state.

Unfortunately, it is \textup{EXPTIME}-complete to find such an optimal
policy in general~\cite{littman:etal:jair-98}. Furthermore, recent
experiments have
shown that, even with very specific restrictions on the action model,
finding an optimal policy for a penetration testing task is feasible
only for small networks of up to 25 hosts~\cite{steinmetz:etal:jair-16}. 
For the sake of scalability and following the lines of Stackelberg Planning~\cite{aaai-stackelberg}, we thus
focus on finding \emph{critical attack paths}, instead of entire
policies.\footnote{Similar approximations have been made in the
  attack-graph literature. Huang et~al.\ \cite{huang:etal:acsac-11},
  e.g., try to identify critical parts of the attack-graph by
  analysing only a fraction thereof, in effect identifying only the
  most probable attacks.}
In a nutshell, a critical attack path is a sequence of actions whose
success probability is maximal. We will also refer to such paths as
\emph{optimal attack plans}, or \emph{optimal attack action
  sequences}.
In contrast to policies, if any action within a critical attack path
does not result in the desired outcome, we consider the attack to have
failed.
Critical attack paths are conservative approximations of optimal
policies, \ie, the success probability of a critical attack path is a
lower bound on the success probability of an optimal policy.

\begin{example}\label{ex:availability}
    Reconsider the outcomes of action $a$ from Example~\ref{ex:acts},
    $O(a)=\{o_\mathit{success},o_\mathit{fail},o_\mathit{crash}\}$.
    Assuming a reasonable set of attacker actions similar to the
    previous examples, no critical path will rely on the outcomes
    $o_\mathit{fail}$
    or
    $o_\mathit{crash}$, as otherwise $a$ would be redundant or
    even counter-productive. 
    Thus the distinction between these two kinds of failures becomes
    unnecessary, which is reflected in the models we generate in
    Section~\ref{sec:model-acquisition} and~\ref{sec:experiments}.
    \begin{full}
    The downside of considering only single paths instead of policies can be
    observed in the following example.
    Consider the case where a second action $a'$ has similar outcomes
    $O(a')=\{o'_\mathit{success},o'_\mathit{fail},o'_\mathit{crash}\}$
    to $a$, but
    $p(o'_\mathit{success}) < p(o_\mathit{success})$
    while
    $p(o'_\mathit{crash})$ is considerably smaller than $p(o_\mathit{crash})$.
    Assuming that $W$ is the only host that can be used to reach $S$
    or $D$, an optimal policy might chose $a'$ in favour of $a$, while
    a critical attack path will insist on $a$.
    \end{full}
\end{example}

Finding possible attacks, \eg, through a penetration testing task as defined
above, is only the first step in securing a network.
Once these are identified, the analyst or the operator need to come up 
with a mitigation plan to mitigate or contain the identified weaknesses. 
This task can be formalized as follows.

\begin{definition}[mitigation-analysis task~\cite{aaai-stackelberg}]
Let $\nprops$ be a set of network propositions, and let $\ptask =
(\aprops, \attacks, \ainit, \goal, \abudget_0)$ be a penetration
testing task. A \emph{$\ptask$ mitigation-analysis task} is a triple
$\mtask = (\ninit, \fixes, \mbudget_0)$ consisting of
\begin{itemize}
\item the \emph{initial network state} $\ninit \subseteq \nprops$,
\item  a finite set of \emph{fix-actions} $\fixes$, and
\item the \emph{mitigation budget} $\mbudget_0 \in \R^+ \cup \{\infty\}$.
\end{itemize}
\end{definition}

The objective in solving such a task --- the defender's objective
---will be to find dominant mitigation strategies within the budget,
\ie, fix-action sequences that reduce the attack probability as much
as possible while spending the same cost. We now specify this in
detail.

Fix-actions encode modifications of the network mitigating attacks
simulated through $\ptask$.

\begin{definition}[fix-actions~\cite{aaai-stackelberg}]
Each fix-action $f \in \fixes$ is a triple $(\pre(f), \post(f), c^\mtask(f))$ of
\emph{precondition} $\pre(f)$ and postcondition $\post(f)$, both 
conjunctions over network proposition literals, and
fix-action cost $c^\mtask(f) \in \R^+$.

We call $f$ \emph{applicable} to a network state $s^\net$ if $\pre(f)$
is satisfied in $s^\net$. The set of applicable $f$ in $s^\net$ is
denoted by $\app(s^\net)$. The result of this application is given by
the state $s^\net\apply{f}$ which contains all propositions with
positive occurrences in $\post(f)$, and all propositions of $s^\net$
whose negation is not contained in $\post(f)$.
\end{definition}

\begin{example}\label{ex:fix-action}
    Removing a vulnerability by, e.g., applying a patch,
    is modelled as 
    a fix-action $f$ with
    $\pre(f)= \vulexists(\cve_W,\allowbreak W,443,\allowbreak\mathsf{tcp},\allowbreak\integrity)$,
    $\post(f)=\neg \pre(f)$
    and cost $1$.

    We can represent adding 
    a firewall 
    between the DMZ and the internet, assuming it was not present
    before,
    as
    a fix-action with
    $\pre(f)=\haclz(\allowbreak \mathsf{internet},\mathsf{dmz},443,\mathsf{tcp}) \land
    \neg \fwapplied(z_2)$,
    $\post(f)= \neg \haclz(\allowbreak \mathsf{internet},\mathsf{dmz},443,\mathsf{tcp}) \land
    \fwapplied(z_2)$
    and cost $100$. 
    It is much cheaper to add a rule to an existing firewall than to
    add a firewall, which can be represented by a similar rule with
    $\fwapplied(z_2)$ 
    instead of
    $\neg\fwapplied(z_2)$ 
    in the precondition, and lower cost.
\end{example}

Note that, in contrast to attacker actions, fix-actions $f$ are deterministic.
A sequence of fix-actions can be applied to a network in order to
lower the success probability of an attacker.

\begin{definition}[mitigation strategy~\cite{aaai-stackelberg}]\label{def:mitigation-strategy} 
    A sequence of fix-actions $\sigma = f_1, \dots, f_n$ is called a
    \emph{mitigation strategy} if it is applicable to the initial
    network state and its application cost is within the available
    mitigation budget\begin{conf}.\end{conf}
    \begin{full}
    , where
\begin{itemize}
\item $f_1, \dots, f_n$ are said to be applicable to a network state $s^\net$
    if $f_1$ is applicable to $s^\net$ 
        and $f_2, \dots, f_n$ are applicable to
        $s^\net\apply{f_1}$.
        The resulting state is denoted $s^\net\apply{f_1, \dots, f_n}$.
\item Applying $f_1, \dots, f_n$ costs
    $c^\mtask(f_1, \dots, f_n) = \sum_{i=1}^{n} c^\mtask(f_i)$. 
\end{itemize}
\end{full}
\end{definition}

To evaluate and compare different mitigation strategies, we consider
their effect on the optimal attack. As discussed in the previous
section, for the sake of scalability we use critical attack paths
(optimal \ie\ maximum-success-probability attack-action sequences) to
gauge this effect, rather than full optimal MDP policies.
As attacker actions in $\ptask$ may contain a precondition on the network state,
changing the network state affects the attacker actions in the state space of
$\ptask$, and consequently the critical attack paths. 
To measure the impact of
a mitigation strategy, we define $p^*(s^\net)$ to be the success probability of
a critical attack path in $s^\net$, or $p^*(s^\net) = 0$ if there is no critical
attack path (and thus there is no way in which the attacker can achieve its
goal).

\begin{definition}[dominance, solution~\cite{aaai-stackelberg}]\label{def:dominance}
Let $\mitigation_1, \mitigation_2$ be two mitigation strategies. $\mitigation_1$
\emph{dominates} $\mitigation_2$ if 
\begin{enumerate}[(i)]
\item $p^*(\ninit\apply{\mitigation_1}) < p^*(\ninit\apply{\mitigation_2})$ and
$c^\mtask(\mitigation_1) \leq c^\mtask(\mitigation_2)$, or
\item $p^*(\ninit\apply{\mitigation_1}) \leq p^*(\ninit\apply{\mitigation_2})$ and
$c^\mtask(\mitigation_1) < c^\mtask(\mitigation_2)$.
\end{enumerate}
The \emph{solution} $\pareto$ to $\mtask$ is the \emph{Pareto
  frontier} of mitigation strategies $\mitigation$: the set of
$\mitigation$ that are not dominated by any other mitigation strategy.
\end{definition}

\begin{full}
In other words, we consider a mitigation strategy $\mitigation_1$
\emph{better} than another one, $\mitigation_2$, if either
$\mitigation_1$ reduces the probability of an successful attack to the
network more, while not imposing a higher cost, or $\mitigation_1$
costs less than $\mitigation_2$ while it lowers the success
probability of an attack at least by the same amount. The solution to
our mitigation-analysis task is the set of \emph{dominant}
(non-dominated) mitigation strategies.
\end{full}

\section{Practical Model Acquisition}\label{sec:model-acquisition}

  In this section, we describe a highly automated approach to acquire
  network models in practice, demonstrating
  our method to be readily applicable.
  Our workflow follows the same idea, but in addition we incorporate 
  possible mitigation actions described in a concise and general
  schema.
  Moreover, our formalism considers the
  probabilistic/uncertain nature of exploits.

\subsection{Workflow}

\begin{figure*}[h] 
\centering
    \begin{tikzpicture}[node distance=1cm, auto,>=latex', thick]
    \path[->] node[tool] (nvd) {NVD}
              node[icon, right=3cm of nvd.center] (vxmli) {\includegraphics[width=1cm]{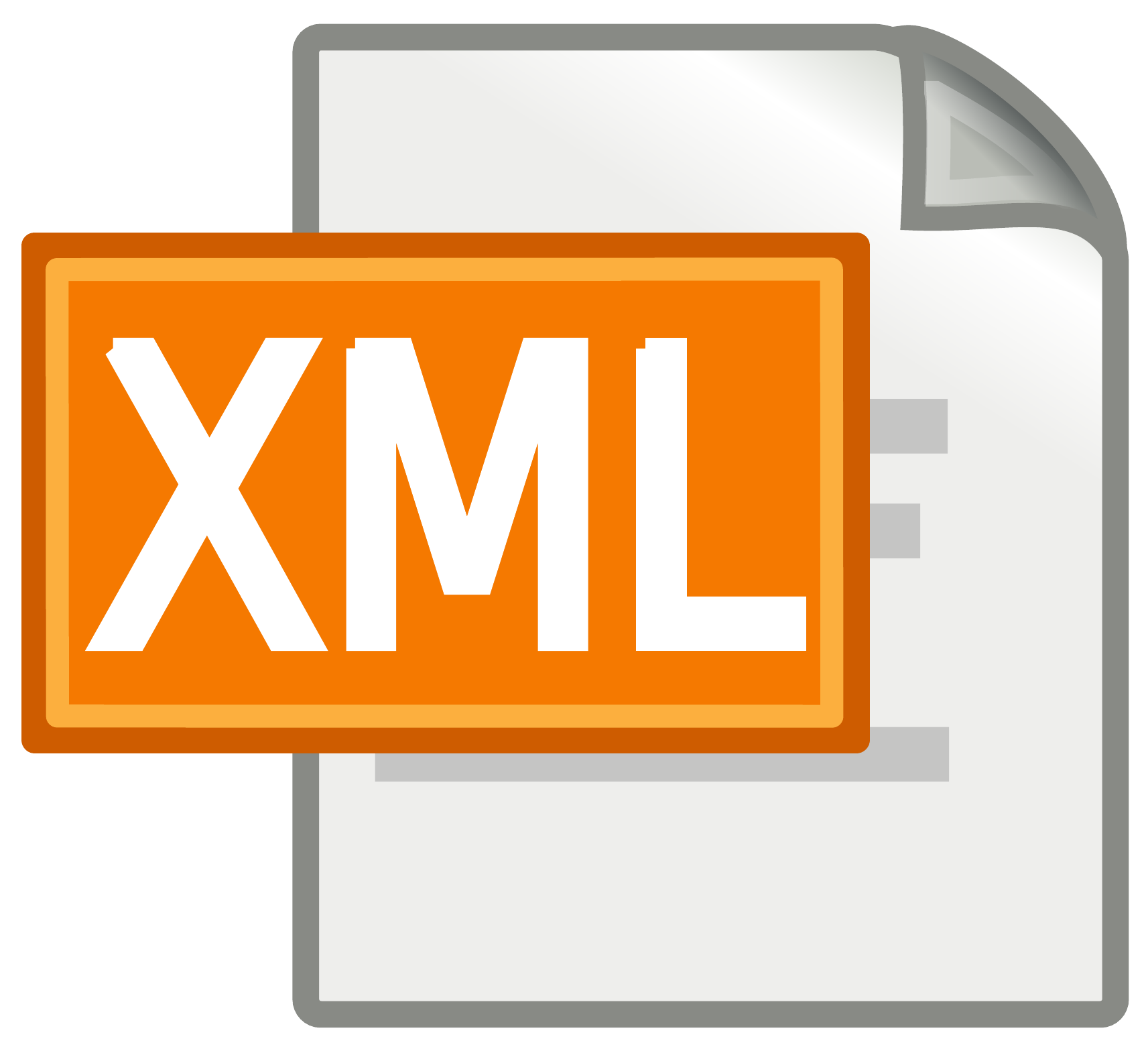}}
              node[file, right=-3mm of vxmli.east, anchor=west] (vxml) {vulnerability.xml}
                  (nvd) edge node {provides} (vxmli);
    \path[->] node[tool, below of=nvd] (nessus) {Nessus}
              node[icon, right=3cm of nessus.center] (rxmli) {\includegraphics[width=1cm]{gfx/xml-icon.pdf}}
              node[file, right=-3mm of rxmli.east, anchor=west ] (rxml) {report.xml}
                  (nessus) edge node {creates} (rxmli);
        \path[->] node[below = 0.5cm  of nessus] (network) {\includegraphics[height=1.4cm]{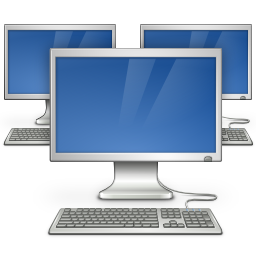}}
              node[below = 0.5cm of network] (user)
              {\includegraphics[height=1cm]{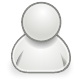}}
                  (nessus) edge [left] node {scans} (network)
                  (user) edge node {describes} (network)
              node[icon, right=3cm of user.center] (fixi) {\includegraphics[width=1cm]{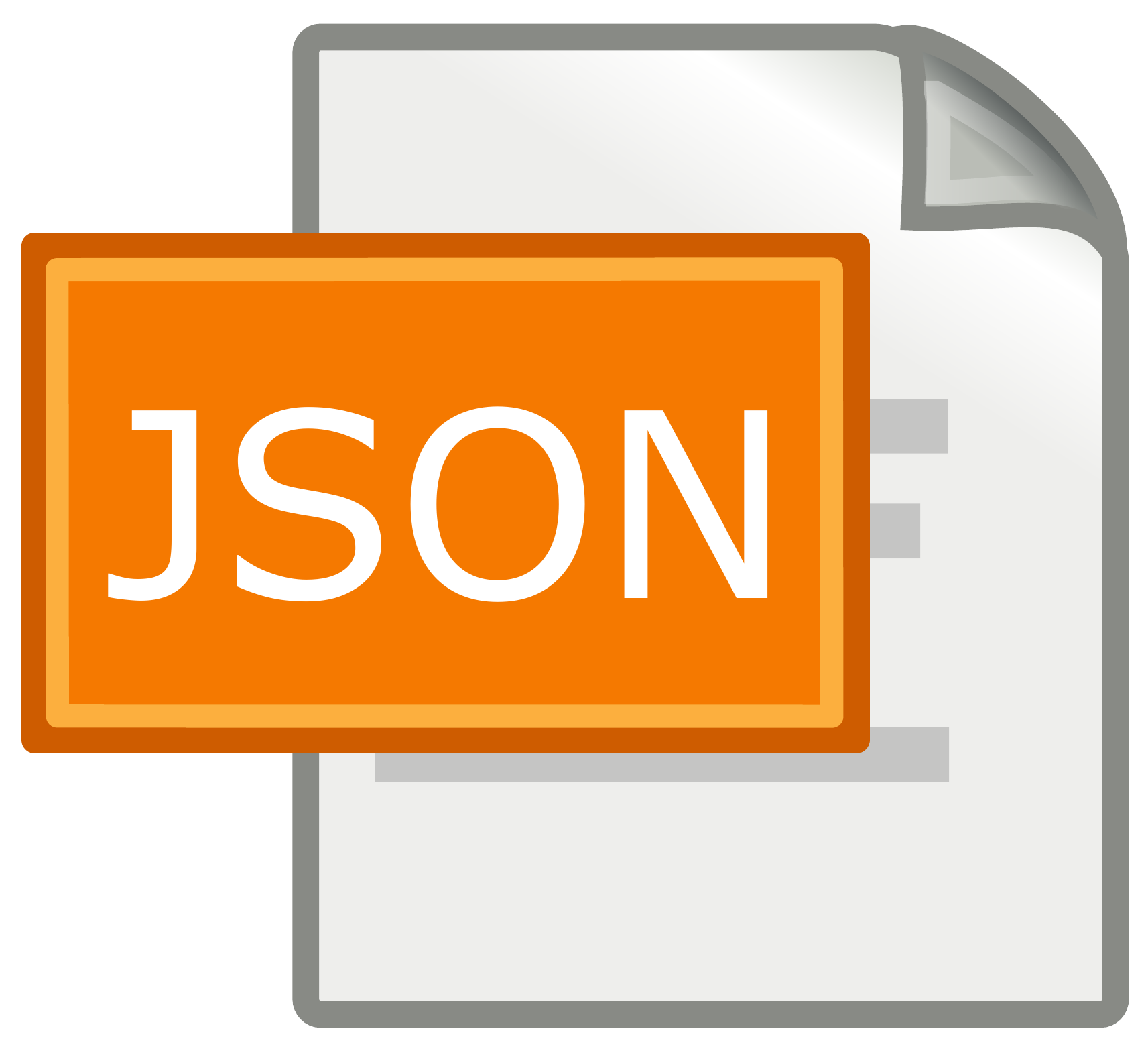}}
              node[file, right=-3mm of fixi.east, anchor=west] (fix) {fixes.json}
                  (user) edge node {creates} (fixi)
              node[icon, above=1.3cm of fixi.west, anchor=west] (topi) {\includegraphics[width=1cm]{gfx/Text-json.pdf}}
              node[file, right=-3mm of topi.east, anchor=west] (top) {topology.json$^*$}
                  (user) edge node {creates} (topi)
              node[icon, below=1.3cm of fixi.west, anchor=west] (acti) {\includegraphics[width=1cm]{gfx/Text-json.pdf}}
              node[file, right=-3mm of acti.east, anchor=west] (act) {actions.json$^*$}
                  (user) edge node {creates} (acti)
                  ;
    \path[->] node[tool, right=2cm of top] (problem) {problem generator}
                  (vxml) edge (problem.north)
                  (rxml) edge (problem.north west)
                  (top) edge (problem)
                  (fix) edge (problem.south west)
                  (act) edge (problem.south);
    \path[->] node[tool, right=1cm of problem.east, anchor=west] (whatif) {what-if analysis}
                  (problem) edge (whatif);
    \path[->] node[tool, below=1.3cm of whatif.center, anchor=center] (result) {result}
                  (whatif) edge (result)
                  (result) edge node [above] {refinement} (fix)
                  ;

        \draw[->]       (result) |- node [above, near end] {refinement} (act);

    \path node[right=2mm of network.east] (networktext) {Network};
    \path node[below=0mm of user.south] (usertext) {User};

\end{tikzpicture}
      \vspace{-0.5cm}
\caption{Workflow for model acquisition via network scanning, assuming
    a fixed attacker and mitigation budget. User
    input marked with $^*$ can be empty. The file \texttt{topology.json}
    can be left empty, in which case an open network is assumed.}
      \vspace{-0.5cm}
\label{fig:workflow}
\end{figure*} 

This section describes the workflow for model acquisition and
refinement via network scanning depicted in Figure~\ref{fig:workflow}.
In the first step, the user scans a network using the Nessus tool,
resulting in a report file.
\begin{full}
          Our current
          toolchain supports only network-wide scans. Nessus, as well as
          several OVAL interpreters~\cite{baker:2011:oval} supports
          host-wise scans, which can be gathered centrally. This would
          give much more precise results, which can be translated in
          a similar way.
\end{full}
The user optionally describes the network topology in a JSON formatted
topology file and sets the hosts that are initially assumed under
adversarial control.\footnote{%
In practice, penetration testers have access to firewall rules in
machine-readable formats (e.g., Cisco, juniper), which can be used to
create this file automatically.}
If this file is not given, we assume all hosts are interconnected
w.r.t.\ every port that appears in the Nessus report.
The user specifies the fixes the analysis should consider. Initially,
this list is (automatically) populated by considering all known
patches and a generic firewall rule that considers adding a firewall
at all possible positions in the network, for the cost of five
patches.
The cost can be refined step by step, and patches that are not
applicable, e.g., because of software incompatibilities, can be
deleted from this file.
The user can also refine the attacker budget and the
mitigation budget. Initially, the attacker budget gives
the number of exploits the attacker may use, as all
exploits are assigned unit cost.  
With this information, the analysis gives a Pareto-optimal set of
mitigation strategies within the given budget. 
After observing the fix-actions, the user may refine the
fix-actions, as adopting some patches might be more
expensive than others (which can be reflected in the
associated mitigation costs), or some firewalls proposed 
might be too
restrictive (which can be
reflected by instantiating the firewall rule).

\subsection{Network Topology and Vulnerabilities}

Like in Example~\ref{ex:network-topology}, the network topology is
given in terms of network predicates $\subnet(z,h) \in \ninit$ for every
host $h$ in subnet $z$, $\haclz(z_1,z_2,\allowbreak \port,\proto)\in\ninit$ for
every $z_1$, from where all hosts in $z_2$ are reachable via
$(\port,\proto)$, which are derived from a JSON file, to allow for
easy manual adjustment.

We translate the Nessus report to a set of network predicates
$\vulexists(\allowbreak \cve,h,\port,\proto,\mathit{type})\in\ninit$ for CVE
$\cve$ affecting $h$ on $(\port,\proto)$, with effect on
$\mathit{type}\in\{ \confidentiality, \integrity, \availability \}$,
and an attack-actions $a$ for each $z_1$, $h_1$ in the universe of
subnets and hosts, and $h_2=h$, such that
    \begin{align*}
        \pre^\net(a) =  &
    \subnet(z_1,h_1)\land
     \subnet(z_2,h_2) \\
        & \land \haclz(z_1,z_2,\port,\proto)
        \\
        & \land \vulexists(\cve,h_2,\port,\proto,\mathit{type}),
    \end{align*}
and $O(a)=\{o_\mathit{success},o_\mathit{fail}\}$.
The value of $\mathit{type}$ is determined from the U.S. government
repository of standards based vulnerability management data, short
NVD\@.
As discussed in Example~\ref{ex:availability},
the future availability of a host is
disregarded by critical path analysis. Furthermore, the NVD does not
provide data on potential side effects in case of failure. Thus, we
assume all hosts in the network to be available throughout the attack.

We handle the success probability different from
Example~\ref{ex:acts} by encoding it into the precondition,
so an action with matching probability is chosen. More precisely,
for all $z_1$, $h_1$ in the universe of subnets and hosts, 
and $p'$ in the universe of probabilities,
and $h_2=h$,
there is an action
$a$
with 
    \begin{align*}
        \pre^\net(a) =  &
    \subnet(z_1,h_1)\land
     \subnet(z_2,h_2) \\
        & \land \haclz(z_1,z_2,\port,\proto)
        \\
        & \land \vulexists(\cve,h_2,\port,\proto,\mathit{type},p'),
    \end{align*}
and $O(a)=\{o_\mathit{success},o_\mathit{fail}\}$,
with success probability
$p(O_\mathit{success})=p'$
and
$p(O_\mathit{fail})=1-p'$,
$\post(o_\mathit{fail})=\top$.
As $a$ can only be applied if $p=p'$, this implies
$p(o_\mathit{success})=p$ for $o_\mathit{success}$ the success outcome
of a matching action. The matching action is uniquely determined, as
in any reachable network state, there is at most one proposition
$\vulexists(\cve,h,\port,\proto,\mathit{type},p)$ for any given
$\cve$, $h$, $\port$, $\proto$ and $\mathit{type}$.

Today, the NVD does not provide data on how vulnerabilities may
impact components other than the vulnerable component, e.g., in case
of a privilege escalation. 
Such escalations are typically filed with
$\mathit{type}=\integrity$. 
Hence 
we identify this vulnerability with
a privilege escalation.
\begin{full}
The latest version 3 of the CVSS standard, released in June 2015,
provides a new metric \texttt{in\_scope}
to specifically designate such vulnerabilities. 
While this metric is still not 
specific enough to accurately describe propagation, 
it at least avoids 
this drastic
over-approximation. As of now, all vulnerability feeds provided by the NVD
are classified using CVSSv2, hence we hope for quick adoption of the
new standard.
\end{full}
Consequently, and as opposed to Example~\ref{ex:acts},
$\pre^\attacks(a)=\compromised(h,\integrity)$,
and $\post(o_\mathit{success})=\compromised(h,\mathit{type})$.
CVSSv2 specifies one of three access vectors:
`local', which we ignore altogether,
`adjacent network', which models attacks that can only be mounted
within the same subnet and typically pertain to the network layer, and
`network', which can be mounted from a different network.
The second differs from the third in that the precondition requires
$z_1$ and $z_2$ to be equal.

We assign probabilities according to 
the `access complexity' metric, which combines 
the probability of finding an exploitable configuration,
the probability of a probabilistic exploit to succeed,
and
the skill required to mount the attack into either `low', `medium' or
`high'.
This is translated into a probability $p$
of $0.2$, $0.5$, or
$0.8$, respectively.
Thus
$p(O_\mathit{success})=p'$
and
$p(O_\mathit{fail})=1-p'$,
where
$\post(o_\mathit{fail})=\top$.
The action cost $c(a)$ is set to $1$.
A separate input file permits the user to refine both action cost
and outcome probability of $o_\mathit{success}$ to reflect
assumptions about the skill of the adversary and
prior knowledge about the software configurations in the network.

\subsection{Threat Model}

The network configuration file defines subnets that are
initially under attacker control, in which case
$\compromised(h,\integrity)\in\ainit$,
and
subnets which the attacker aims to compromise, in which case the goal
condition is
\[\bigwedge_\text{$(z,\mathit{type})$ marked as target in \texttt{topology.json}} \targetundercontrol(z,\mathit{type}).\]
%
%
Additional artificial actions permit deriving
$\targetundercontrol(z,\mathit{type})$ whenever
$\compromised(h,\mathit{type})\land \subnet(z,h)$.

\subsection{Mitigation Model}

Our formalism supports a wide range of fix-actions, but to facilitate
its use, we provide three schemas, which we instantiate to a larger
number of actions.

\subsubsection*{Fix schema}

The fix schema models the application of existing patches, 
the development of missing patches
and the implementation of local workarounds, e.g.,
application-level firewalls that protect systems from malicious
traffic which are otherwise not fixable. 
The user specifies the CVE, host and port/protocol the fix applies to.
Any of these may be a wild card *, in which case all matching fix
actions of the form described in Example~\ref{ex:fix-action} are
generated. The schema also includes the new probability assigned
(which can be 0 to delete these actions) and an initial cost, which is
applied the first time a fix-action instantiated from this schema is
used, and normal cost which are applied for each subsequent use.
Thus, the expensive development of a patch (high initial cost, low
normal cost) can be compared with local workarounds that have higher
marginal cost.
The wild cards may be used to model available patches that apply to all
hosts, as well as generic local workarounds that apply to any host, as
a first approximation for the initial model.

Non-zero probabilities may be used to model counter-measures which
lower the success probability, but cannot remove it completely, e.g.,
address space layout randomisation.
We employ a slightly indirect encoding to accommodate this case,
adding additional attack-action copies for the lowered probability.
The network state predicate determines uniquely which attack-action
among these applies. The generated fix-action modifies the network
state predicate accordingly.

\subsubsection*{Firewall schemata}

There are two firewall schemas, one for firewalls between subnets,
one for host-wise packet filtering. The former is defined by source
and destination subnet along with port and protocol. Similar to the
fix schema, any of the value may be specified, or left open as a wild
card *, in which case a fix-action similar to the firewall fix in
Example~\ref{ex:fix-action} is instantiated for every match. In
addition, initial costs and cost for each subsequent application can
be specified, in order to account for the fact that installing
a firewall is more expensive than adding rules. 
The second firewall schema permits a similar treatment per host
instead of subnets, which corresponds to local packet filtering rules.

\section{Experiments}\label{sec:experiments}

\begin{figure*}[th]
\setlength{\tabcolsep}{0pt}
\begin{tabular}{ccc}
\includegraphics[height=3.55cm]{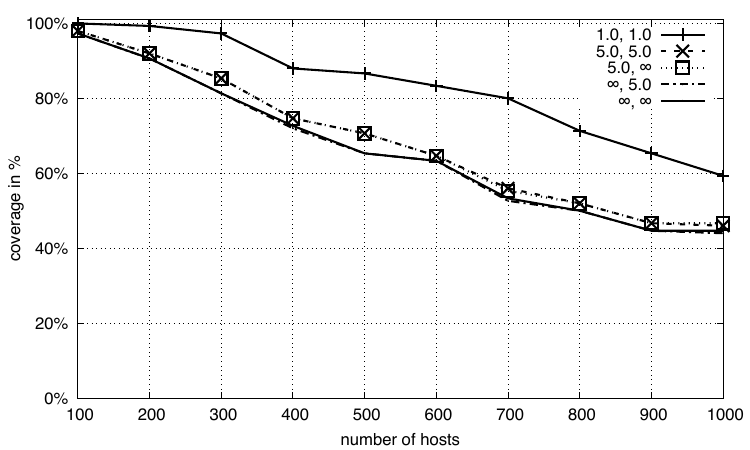}
&
\includegraphics[height=3.55cm]{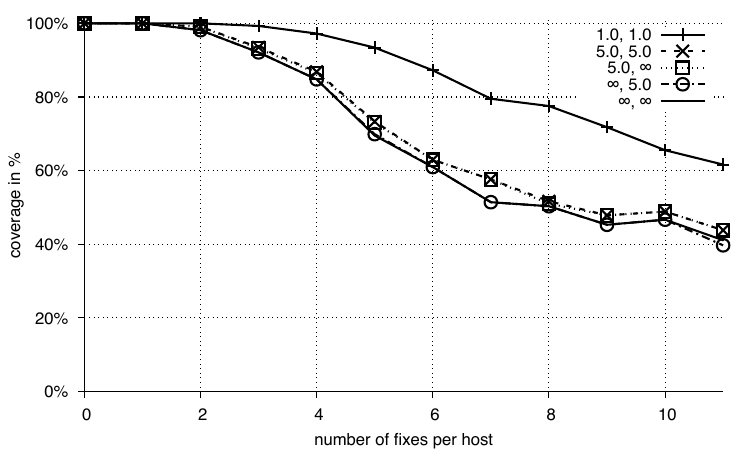}
&
\includegraphics[height=3.55cm]{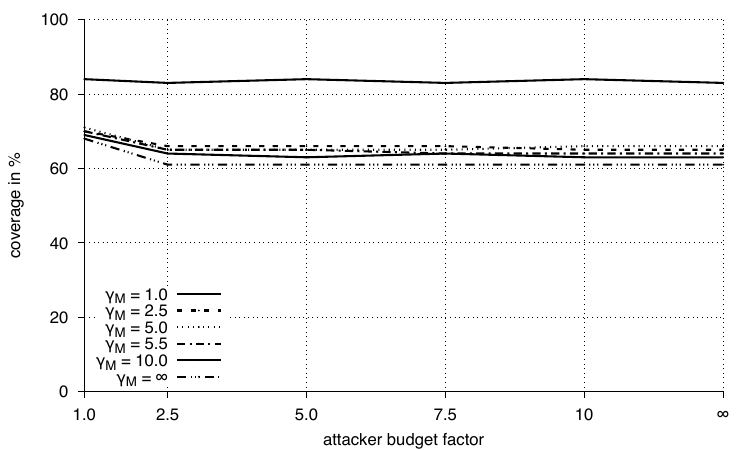}
\\
(a) & (b) & (c)
\end{tabular}
  \vspace{-0.5cm}
\caption{\% instances solved within time (memory) limits. (a) for scaling number
of hosts and $5$ fix-actions per host on average, (b) for scaling number of
fix-actions per hosts, but fixing $H=500$, and (c) scaling budgets, but fixing
the number of hosts to $500$ and fixing the number of fixes to $5$ per host. \vspace{-2em}}%
\label{fig:coverage-by-number-of-fixes-per-hosts-}%
\label{fig:coverage-percentage-of-instance-that-terminated-within-}%
\label{fig:coverage-depending-on-attacker-budget-and-mitigation-budget-with-}
  \vspace{-0.0cm}
\end{figure*}

It is easy to see that Stackelberg planning is PSPACE-hard.
We hence explore the space of problems in which Stackelberg planning
performs well enough to be useful. 
To provide an intuitive account of this space in terms of the network
to be scanned, we
we created a problem generator
that produces network topologies and host configurations based on
known vulnerabilities. 
This facilitates the performance evaluation
of our mitigation analysis
algorithm
w.r.t.\ 
the number of hosts, fix actions and any combination of attacker
and mitigation budget.
For details to the generator, we refer the reader to 
\begin{conf}
the long version of this paper~\cite{whatif-full}.
\end{conf}
\begin{full}
Appendix~\ref{sec:generator}.
\end{full}

We evaluate our model using Speicher et.\, al's Stackelberg planning
algorithm~\cite{aaai-stackelberg} which was implemented on top of the FD
planning tool~\cite{helmert:jair-06}. Our experiments were conducted
on a cluster of Intel Xeon E5-2660 machines running at 2.20 GHz. We
terminated a run if the Pareto frontier was not found within 30
minutes, or the process required more than 4 GB of memory during
execution.

In our evaluation we focus on coverage values, i.e.\ the number of instances that
could be solved within the time (memory) limits. We investigate how coverage is
affected by (1) scaling the network size, (2) scaling the number of fix actions,
and (3) the mitigation budget, respectively the attacker budget. 

The budgets are computed as follows. In a precomputation step, we compute the
minimal attacker budget $\minabudget$ that is required for non-zero success
probability $p^*(\ninit)$. The minimal mitigation budget $\minmbudget$ is then
set to the minimal budget required to lower the attacker success probability
with initial attacker budget $\abudget_0 = \minabudget$. We experimented with
budget values relative to those minimal budget values, resulting from scaling
them by factors out of $\{1, 2.5, 5, 7.5, 10, \infty\}$. We denote
$\mbudgetfactor$ the factor which is used to scale the mitigation budget, and
vice versus $\abudgetfactor$ the factor for the attacker budget.

In Figure~\ref{fig:coverage-percentage-of-instance-that-terminated-within-}(a),
we observe that the algorithm provides
reasonable coverage $>50\%$ for up to $800$ hosts, when considering
on average
5 vulnerabilities and 5 fix-actions per host.
Unless both the attacker and the mitigation budget are scaled to $1$
(relative to \mbudgetfactor\ and \abudgetfactor, respectively),
this result is relatively independent from the budget.
One explanation why it is independent from the budget is that there is no huge difference between factors $5$ and $\infty$ in the sense that the attacker cannot find more or better critical paths and the defender cannot find more interesting fix action sequences because of the infinite budgets.
In the case that both are scaled to $1$, the searches for critical paths
and fix actions sequences are vastly simplified.
Hence the overall coverage is better.

Note that the number of fix actions scales linearly with the number of hosts,
which in the worst case, i.e., when all sequences need to be regarded, leads to
an exponential blowup.

In Figure~\ref{fig:coverage-by-number-of-fixes-per-hosts-}(b), we have fixed the
number of hosts to $500$, but varied the number of fixes that apply per host by
scaling $\lambda_F$ in integer steps from $0$ to $10$, which controls the
expected value of patch fixes generated per host.
We then plotted the coverage with the total number of fixes, i.e., the number of
firewall fixes and patch fixes actually generated.
We tested $50$ samples per value of $\lambda_F$ and attacker/mitigation budget. We
cut of at above 11 fixes per host, where we had too few data points. We furthermore applied a sliding average with a window size of
  $1$ to smoothen the results, as the total number of actual fixes varies for
a given $\lambda_F$.
Similar to
Figure~\ref{fig:coverage-percentage-of-instance-that-terminated-within-}(a), the
influence of the attacker and mitigation budget is less than expected, except for
the extreme case where both are set to their minimal values.
The results suggest that the mitigation analysis is reliable up to a number of
4 fixes per hosts, but up to 16 fixes per host, there is still a decent chance
for termination.

Figure~\ref{fig:coverage-percentage-of-instance-that-terminated-within-}(c)
compares the impact of the mitigation- and attacker-budget factors
$\mbudgetfactor,\abudgetfactor \in \{1,
2.5, 5, 7.5, 10, \infty \}$.
The overall picture supports our previous observations.  The attacker budget has
almost no influence on the performance of the algorithm. 
This, however, is somewhat surprising given that the attacker budget not only
affects the penetration testing task itself, but also influences the
mitigation-analysis. Larger attacker budgets in principle allow for more
attacks, imposing the requirement to consider more expensive mitigation
strategies.
It will be interesting to explore this effect, or lack thereof, on real-life
networks.

In contrast, the
algorithm behaves much more sensitive to changes in the mitigation budget.
Especially in the step fom $\mbudgetfactor=1.0$ to $\mbudgetfactor=2.5$,
coverage decreases significantly (almost 20 percentage points regardless of the
attacker budget value). 
This can be explained by the effect of the increased mitigation budget
on the search space. However,
further increasing the mitigation budget has a less
severe effect.
\begin{full}
Again, we attribute this to the problems we generate: 
In many cases, the mitigation strategy that results in the minimal
possible attacker success probability is cheaper than the mitigation
budget resulting from $\mbudgetfactor=2.5$.
In almost half of the
instances solved for $\mbudgetfactor=2.5$, this minimal attacker success
probability turned out to be $0$. In these cases specifically, 
the mitigation analysis can readily prune mitigation strategies with higher
costs,
even if
more mitigation budget is available,
as the Stackelberg planning algorithm maintains the current bound for the cost of lowering the
attacker probability to zero.
\end{full}

\section{Conclusion \& Future work}

The mitigation analysis method presented in this work is the first of
its kind and provides a semantically clear and thorough methodology
for analysing mitigation strategies.
We leverage the fact that network attackers can be simulated, and
hence strategies for mitigation can be compared
before being implemented.
We have presented a highly automated modelling approach
along with an iterative
workflow.
Based on a detailed network and configuration model, we demonstrated
the feasibility of the approach and scalability of the algorithm.

Two major ongoing and future lines of work arise from this
contribution, pertaining to more effective algorithms, and to the
practical acquisition of more refined models.
Regarding effective algorithms, the major challenge lies in the
effective computation of the Pareto frontier. As of now, this stands
and falls with the speed with which a first good solution --- a cheap
fix-action sequence reducing attacker success probability to a small
value --- is found.
\begin{full}
In case that happens quickly, our pruning methods
and thus the search become highly effective; in case it does not
happen quickly, the search often becomes prohibitively enumerative and
exhausts our 30 minute time limit. In other words, the search may, or
may not, ``get lucky''. What is missing, then, is effective
\emph{search guidance} towards good solutions, making it more likely
to ``get lucky''.
This is exactly the mission statement of heuristic functions in AI
heuristic search procedures. The key difference is that these
procedures address, not a move-countermove situation as in fix-action
sequence search, but single-player (just ``move'') situations (like at
our attack-planning level, where as mentioned we are already using
these procedures). This necessitates the extension of the heuristic
function paradigm --- solving a \emph{relaxed} (simplified) version of
the problem, delivering relaxed solution cost as a lower bound on real
solution cost --- to move-countermove situations. This is a
far-reaching topic, relevant not only to our research here but to AI
at large, that to our knowledge remains entirely
unexplored.\footnote{Game-state evaluation mechanisms are of course
  widely used in game-playing, yet based on weighing (manually or
  automatically derived) state features, not on a relaxation paradigm
  automatically derived from the state model.} Notions of
move-countermove relaxation are required, presumably
over-approximating the defender's side while under-approximating the
attacker's side of the game, and heuristic functions need to be
developed that tackle the inherent min/max nature of the combined
approximations without spending too much computational effort. For our
concrete scenario here, one promising initial idea is to fix the
attacker's side to the current optimal critical attack path, and
setting the defender's objective --- inside the heuristic function
over-approximation --- as reducing the success probability of that
critical attack path as much as possible, while minimizing the
summed-up fix-action cost. This results in an estimation of fix-action
quality, which should be highly effective in guiding the fix-action
level of the search towards good solutions quickly.
\end{full}
\begin{conf}
    Finding these good solutions quickly is precisely the mission
    statement of heuristic functions in AI heuristic search
    procedures, which typically operate by solving a \emph{relaxed}
    (simplified) version of the problem to deliver lower bounds.
    The key difficulty is the move-countermove
    pattern in Stackelberg planning, which requires a new
    understanding of what a relaxation ought to achieve in this
    setting.
\end{conf}

Regarding the practical acquisition of more refined models, 
the quality of the results of our analyses of course hinges on the
accuracy of the input model.
In practice, there is a trade-off between the accuracy of the model,
and the degree of automation vs.\ manual effort with which the model
is created.
\begin{full}
This is partially due to the fact that vulnerabilities are often
discovered in the process of pentesting, which a simulation cannot
reproduce.
(Although potential zero-day exploits can in principle be
modeled in our framework as a particular form of attack-actions, that
exist only with a given probability.)
It is also due to the fact that current vulnerabilities lack
necessary information to derive these models automatically.
There are two factors to the latter. 
\end{full}
First, economically, a more detailed machine-readable description of
vulnerabilities cost money, hence there needs to be
an incentive to provide this data. 
The successful commercial use of simulated pentesting at Core Security
shows that there is money to be made with
fine-grained  vulnerability data.
We hope that mitigation analysis methods such as ours will be adopted
and provide further incentives, as centralised knowledge about the
nature of vulnerabilities can be used to improve analysis and hence
lower mitigation cost. Declarative descriptions like OVAL are
well-suited to this end.

Second, conceptually, the transitivity in network attacks is
not understood well enough.
Due to the lack of additional information, we assume that integrity
violations allow for full host compromise, which is an
over-approximation.
While CVSSv3 provides a metric distinguishing attacks that switch scope,
it is unclear how exactly this could be of use, as the scope might
pertain to user privileges within a service, sandboxes, system users,
dom0-privileges etc.
A formal model for privilege escalation could be used to describe the
effect if a vulnerability in an abstract manner that can be
instantiated into a concrete outcome once an actual software
configuration is given and form the basis for the automated 
acquisition of realistic network models.

\bibliographystyle{ACM-Reference-Format}
\bibliography{abbreviations,biblio,crossref}

\begin{full}
\appendix
\section{Scenario Generator}

\label{sec:generator}

\begin{figure}[h] 
\centering
    \begin{tikzpicture}[node distance=1.5cm, auto,>=latex', thick]
        \path[->] 
              node[dist] (uni) {$\rUnif$}
              node[var, left of= uni] (v) {$\NVD$}
              node[dist, below of= uni] (V) {$\rVuln$}
              node[var, left of= V] (alphav) {$\alpha_V$}
              node[dist, below of= V] (C) {$\rConfs$}
              node[dist, right of = V] (N) {$\rLength$}
              node[var, right of = N] (l) {$\lambda$}
              node[dist, below of= C] (H) {$\rHosts$}
              node[dist, below of=H] (hi) {$h_i$}
              node[var, left of= H] (alphah) {$\alpha_H$}
              node[below=1em of hi.south] (label) {\scriptsize $i=1,\ldots,n$}
                  (v) edge (uni)
                  (uni) edge (V)
                  (alphav) edge (V)
                  (V) edge (C)
                  (l) edge (N)
                  (N) edge (C)
                  (alphah) edge (H)
                  (C) edge (H)
                  (H) edge (hi)
                  ;

        \node[fit=(hi)(label), draw, inner sep=3mm] (square) {};

        \node[right=1em of uni.east] {(uniform distr. over
        vulnerabilities)};
        \node[right=1em of C.east] {(base distr. over
        configurations)};
        \node[right=1em of H.east] {(distribution of hosts)};
        \node[right=1em of square.east, align=left, text width=5cm]
        { (host configurations assigned to network nodes)};
        
\end{tikzpicture}
\caption{Probabilistic graphical model for the distribution of
    configurations within the network.}
\label{fig:pgm}
\end{figure}
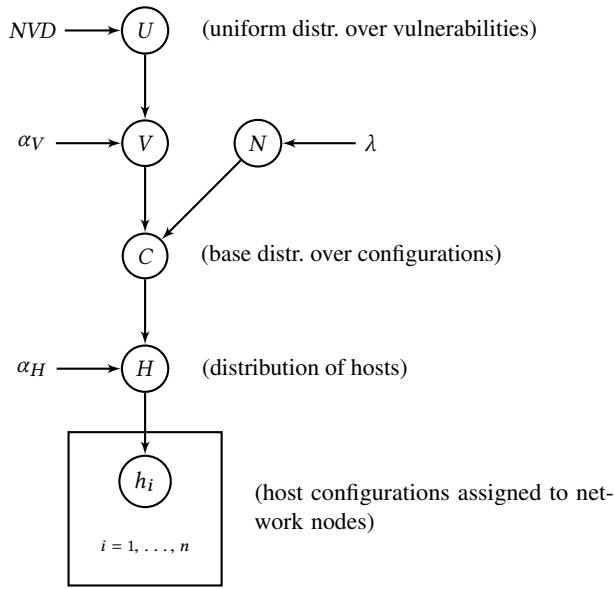 

The problems we generate are modelled exactly as described in
Section~\ref{sec:model-acquisition}. We hence describe the scenario generation
in terms of the topology, \ie, subnet relations defined by the network
proposition $\subnet$ and connections described by the network proposition
$\haclz$, and the assignment of configurations, i.e., the network proposition
$\vulexists$ and corresponding actions.

\subsubsection*{Topology} The network topology generation follows previous works
on network penetration testing task generators \cite{sarraute:etal:aaai-12}.
Similar to the running example, we generate networks which are partitioned into
four zones: users, DMZ, sensitive and internet. The internet consists of only
a single host which is initially under adverserial control, and which is
connected to the DMZ zone. The DMZ and the sensitive zone each constitute
a subnet of hosts, both subnets being connected to each other. The user zone is
an hierarchy, tree, of subnets, where every subnet is connected to its parent and
the sensitive part. Additionally, the root subnet of the user zone is also
connected to the DMZ zone. A firewall is placed on every connection between
subnets. While the firewalls inside the user zone are initially empty, \ie, they
do not block anything, the firewalls located on connections between two
different zones only allow traffic over a fraction of ports. The ports blocked
initially are selected randomly.  The size of the generated networks is scaled
through parameter $H$ determining the overall number of hosts in the network. To
distribute $H$ hosts to the different zones, we add for each 40 hosts one to
DMZ, one to the sensitive zone, and the remaining to the user zone (\cf
\cite{sarraute:etal:aaai-12}).

\subsubsection*{Configurations}

Now we come to the assignment of configurations, i.e., set of vulnerabilities to
hosts.
In many corporate networks, host configurations are standardized, e.g.,
workstations have equivalent configurations or each machine within a cluster is
alike. To this end, we model the distribution of the totality of hosts used in
the network by means of a (nested) \emph{Dirichlet process}. 

Depending on the concentration parameter $\alpha_H$, the $i$th host $H_i$ is,
with probability $\alpha_H/(\alpha_H+n-1)$, drawn freshly from the distribution
of configurations $\rConfs$, which we will explain in the followup, or otherwise
uniformly chosen among all previous $H_j$, $j=1,\ldots,i-1$.
Formally, $\rHosts \mid \rConfs \sim \mathrm{DP}(\alpha_H,\rConfs)$.
Configurations are drawn using the following process.
First, the number of vulnerabilities a configuration has in total is drawn
according to a Poisson distribution, $\rLength \sim \mathrm{Pois}(\lambda_V)$.
This number determines how many vulnerabilities are drawn in the next step by
means of a second Dirichlet process.
This models the fact that the software which is used in the same company tends
to repeat across configurations.
The base distribution over which the Dirichlet process chooses vulnerabilities
is the uniform distribution over the set of vulnerabilities in our database
$\NVD$, i.e., $\rVuln \sim \mathrm{DP}(\alpha_V,\rUnif_\NVD)$.

A configuration $\rConfs$ is now chosen by drawing 
$n$ from $\rLength$ and then drawing $n$ samples from $\rVuln$, i.e.,
\[
    \Pr [
        \rConfs = (c_1,\ldots,c_n)
    ]  = \Pr[ \rVuln=n, c_1=\rVuln_1, \ldots,c_n=\rVuln_n  \ ].
    \]
where $\rVuln_1,\ldots,\rVuln_n\sim \rVuln$.
Observe that $\rConfs$ are conditionally independent given 
$\mathrm{DP}(\alpha_V,U_\NVD)$, hence $\rConfs$ may define the base
distribution for the above mentioned Dirichlet process.
Now the configurations are drawn from the distribution we just
described, i.e., $h_1,\ldots,h_n \sim \rConfs$.

We are thus able to control the homogeneity of the network, as well
as, indirectly, via $\alpha_V$, the homogeneity of the software
configurations used overall. For example, if $\alpha_H$ is low but
$\alpha_V$ is high, many configurations are equal, but if they are
not, they are likely to not have much intersection (provided
$\mathit{NVD}$ is large enough). If $\alpha_V$ is low but $\alpha_H$
is high, many vulnerabilities reappear in different host
configurations.

\subsubsection*{Mitigation model} Akin to Section~\ref{sec:model-acquisition},
we consider two different types of fix-actions: closing open ports by adding
rules to firewalls, and closing vulnerabilities through applying known patches.
For the former, we generate for each subnet and each port available in this
subnet a fix-action $f$ that in effect blocks all connections to this subnet
over this port, by negating the corresponding $\haclz$ propositions.  Such
fix-actions are always generated for all user subnets. DMZ and sensitive
conceptionally do not allow closing all ports, as some ports must remain opened
for services running in those subnets. Hence, for DMZ and the sensitive zone, we
randomly select a subset of open ports which must not be locked out through
firewall rules, and firewall fix-actions are then only generated for the
remaining ports.  Patch fix-actions are drawn from the set of possible patches
described inside the OVAL database that is provided from Center for Internet
Security.\footnote{\url{https://oval.cisecurity.org/repository}} In OVAL, each
patch is described in terms of an unique identifier, human readable metadata,
and a list of vulnerabilties that are closed through the application of this
patch. We assign patch actions to each generated configuration. Similar to
before, first, the number of patches a configuration has in total is drawn
according to a Poisson distribution, $\rLength_F \sim \mathrm{Pois}(\lambda_F)$.
For each configuration, first the actual number of patches $n_F$ is sampled from
$\rLength_F$, and then $n_F$ patches are drawn uniformly from the set of patches
given by OVAL which affect at least one vulnerability for this configuration.

\end{full}
\end{document}